\documentclass{article}
\usepackage{amssymb}
\usepackage{amsmath}
\usepackage{amstext}
\usepackage{amsfonts}
\numberwithin{equation}{section}
\usepackage[a4paper, top=3cm, bottom=4cm]{geometry}
\usepackage[latin1]{inputenc}
\usepackage{setspace}
\usepackage{fancyhdr}
\usepackage{slashed}
\usepackage{simplewick}
\usepackage{textcomp}
\usepackage{gensymb} 
\usepackage{xcolor}
\usepackage{rotating}
\usepackage{titlesec}
\usepackage{relsize}
\usepackage{graphicx}

\newcommand{\lie}{\pounds}

\usepackage{epsfig}
\usepackage{dcolumn}
\usepackage{color}
\usepackage{comment}
\usepackage{hyperref}

\def\XXint#1#2#3{{\setbox0=\hbox{$#1{#2#3}{\int}$ }
\vcenter{\hbox{$#2#3$ }}\kern-.56\wd0}}

\newcommand*\xbar[1]{%
  \hbox{%
    \vbox{%
      \hrule height 0.5pt 
      \kern0.5ex
      \hbox{%
        \kern-0.1em
        \ensuremath{#1}%
        \kern-0.1em
      }%
    }%
  }%
}

\definecolor{rosso}{cmyk}{0,1,1,0.4}
\definecolor{rossos}{cmyk}{0,1,1,0.55}
\definecolor{rossoc}{cmyk}{0,1,1,0.2}
\definecolor{blu}{cmyk}{1,1,0,0.3}
\definecolor{blus}{cmyk}{1,1,0,0.6}
\definecolor{bluc}{cmyk}{1,1,0,0.1}
\definecolor{verde}{cmyk}{0.92,0,0.59,0.25}
\definecolor{verdec}{cmyk}{0.92,0,0.59,0.15}
\definecolor{verdes}{cmyk}{0.92,0,0.59,0.7}

\newcommand{\ba}{\begin{eqnarray}}
\newcommand{\ea}{\end{eqnarray}}
\newcommand{\be}{\begin{equation}}
\newcommand{\ee}{\end{equation}}
\newcommand{\bi}{\begin{itemize}}
\newcommand{\ei}{\end{itemize}}
\newcommand{\al}{\alpha}
\newcommand{\bt}{\beta}

\newcommand{\ka}{\kappa}

\newcommand{\sa}{\sigma}

\newcommand{\cH}{{\cal H}}

\newcommand{\cL}{{\cal L}}

\newcommand{\p}{\partial}

\newcommand{\LT}{\left[}
\newcommand{\RT}{\right]}

\begin{document}

\title{Rotating black hole and entropy  for \\modified theories of gravity}
\author{Spyridon Talaganis, Ali Teimouri
\\ \\
 {\it Consortium for Fundamental Physics, Lancaster University,} \\
{\it Lancaster, LA$1$ $4$YB, United Kingdom.}\\
\\
\textit{E-mail}:  s.talaganis@lancaster.ac.uk, a.teimouri@lancaster.ac.uk}
\date{\today}
\maketitle

\begin{abstract}
    In this paper we obtain the entropy of the Kerr black hole for a number of modified theories of gravity. We show that as long as the deviation from Einstein Hilbert term consists  purely of terms involving scalar curvature  and Ricci tensor, the entropy is solely given by the area law. However, the area law will be modified by appropriate corrections when we consider terms involving Riemann tensor explicitly. 
    \end{abstract}


\section{Introduction}
It is well established that the black holes behave as  thermodynamical systems \cite{Bardeen:1973gs}. The first realisation of this fact was made by Hawking, \cite{Hawking:1974sw}.  It is discovered that quantum processes make black holes to emit a thermal flux of particles. As a result, it is possible for a black hole to be in thermal equilibrium with other systems.  We shall recall the thermodynamical laws that govern black holes: The zeroth law states that the horizon of stationary black holes have a constant surface gravity. The first law states that when stationary black holes are being perturbed the change in energy is related to the change of area, angular momentum and the electric charge associated to the black hole. The second law states that, upon satisfying the null energy condition the surface area of the black hole can never decrease. This is the law that was realised by Hawking as the area theorem and showed that black holes radiate. Finally, the third law states that the black hole can not have vanishing surface gravity.

The second law of the black holes' thermodynamics requires an entropy for black holes. It was Hawking and Bekenstein, \cite{Bekenstein:1973ur}, who conjectured that black holes' entropy is proportional to the area of its event horizon divided by Planck length. Perhaps, this can be seen as one of the most striking conjectures in modern physics. Indeed, through Bekenstein bound, \cite{Bekenstein:1972tm},  one can see that the black hole entropy, as described by the area law, is the maximal entropy that can be achieved and this was the main hint that led to the holographic principle, \cite{Susskind:1994vu}.

The black hole entropy can be obtained through number of ways. For instance, Wald \cite{Wald:1993nt} has shown that the entropy for a spherically symmetric and stationary black hole can be obtained by calculating the  N\"oether charge. Equivalently, one can obtain the change in mass and angular momentum by using the Komar formula and subsequently use the definition of the first law of the black holes' thermodynamics to obtain the entropy. Normally, obtaining the entropy for non-rotating black holes is very straightforward. In this case, one uses the  Schwarzschild metric and follow the Wald's approach to calculate the entropy. Also for rotating black holes that are described by Kerr metric one can simply use the Komar integrals to find the mass and angular momentum and finally obtain the entropy. However, when we deviate from Einstein's theory of general relativity obtaining the conserved charge and hence the entropy can be challenging. See these \cite{Peng:2014gha,Barnich:2004uw,Barnich:2003xg,Barnich:2001jy} for advancement in finding the conserved charges. 

The Einstein's theory of general relativity had been very successful in predicting the natural phenomena. However, the theory is sick and is not sufficient to explain many of the fundamental problems. Physicists in the past decades attempted to resolve these problems by modifying the theory of general relativity. For instance, $f(R)$ gravity \cite{Sotiriou:2008rp}  is a generalisation of Einstein-Hilbert action. The most famous type of $f(R)$ gravity was introduced by Starobinsky, \cite{Starobinsky:1980te}, to explain inflation. There are many other modified theories of gravity that attempt to solve the gravitational problems. For partial list of these theories see \cite{Clifton:2011jh}. As various theories of gravity are advancing, it is essential to study different physical aspects of these theories and perhaps examine their viability. So far, many calculations have been done to obtain the entropy of black holes for modified gravitational theories (\textit{e.g.} \cite{Conroy:2015wfa,Conroy:2015nva}). However, in most cases simple solutions were studied, namely those black holes that are described by diagonal metrics, such as Schwarzschild, (A)dS, and etc. This is due to the fact that more complicated metric solutions, such as full Kerr metric, contain off-diagonal terms that make the calculations very challenging. 

In this paper, we are going to briefly review the notion of N\"oether and Komar currents in variational relativity. We show how the two are identical and then we move to calculate the entropy of Kerr black holes for a number of examples, namely $f(R)$ gravity, $f(R,R_{\mu\nu})$ theories where the action can contain higher order curvatures up to Ricci tensor and finally higher derivative gravity. The entropy in each case is obtained by calculating the modified Komar integrals.     

\section{Variational principle, N\"oether and Komar currents }
Variational principle is a powerful tool in physics. Most of the laws in physics are derived by using this rather simple and straightforward method.  Given a gravitational  Lagrangian, $L$ \textbf{}, we can obtain the equations of motion by simply varying the action with respect to the inverse metric, $g^{\mu\nu}$. In short form, this can be done by defining two \textit{covariant momenta}:
\begin{equation}
\pi_{\mu\nu}=\frac{\delta \mathcal{L}}{\delta g^{\mu\nu}}, \quad P^{\mu\nu}=\frac{\delta \mathcal{L}}{\delta R_{\mu\nu}},
\end{equation}
and thus the variation of the Lagrangian would be given by \cite{Fatibene:1998rq}: 
\begin{equation}
\delta \mathcal{L}=\pi_{\mu\nu}\delta g^{\mu\nu}+P^{\mu\nu}\delta R_{\mu\nu}. 
\end{equation}
It is simple to see that in the example of Einstein Hilbert (EH) action,  the first term admits the equations of motion (\textit{i.e.} $\pi_{\mu\nu}=0$) and the second term will be the boundary term. Since we are considering gravitational theories, the general covariance must be preserved at all time. In other words, the Lagrangian, $L$, is covariant with respect to the action under diffeomorphisms of space-time. Infinitesimally, the variation can be expressed as: \begin{equation}
\delta_{\xi}L=d(i_{\xi}L)=\pi_{\mu\nu}\lie_{\xi} g^{\mu\nu}+P^{\mu\nu}\lie_{\xi} R_{\mu\nu},
\end{equation}
where $\delta_{\xi}$ denotes an infinitesimal variation of the gravitational action, $d$ is the exterior derivative, $i_\xi$ is interior derivative of forms along vector field $\xi$ and $\lie_{\xi}$ is the Lie derivative with respect to the vector field. By expanding the Lie derivative of the Ricci tensor, and noting that:
\begin{equation}\label{metricvar}
\delta_{\xi}g_{\alpha\beta}=\pounds_{\xi}g_{\alpha\beta}=\nabla_\alpha\xi_\beta+\nabla_\beta\xi_\alpha,
\end{equation}
 the  \textit{N\"other conserved current} can be obtained.  The way this can be done for the EH\ action is  demonstrated in
Appendix \ref{ehcons} as an example. Furthermore, the conserved \textit{N\"oether current associated to the general covariance of the Einstein- Hilbert action is identical to the generalised Komar current}. This can be seen explicitly in Appendix
\ref{generalisedkomar}. In general, we define the Komar current\footnote{As a check it can be seen that for EH\ action we have, 
$$\label{ehpotential}
P^{\alpha\beta}_{EH}=\sqrt{-g}g^{\alpha\beta},\qquad\mathcal{U}_{EH}=\sqrt{-g}\nabla_{\alpha}\xi^{[\mu}g^{\nu]\alpha}ds_{\mu\nu}
$$
which is exactly the same as what we obtain in Eq. (\ref{noethercurrent}).}as \cite{Fatibene:1998rq}: 
\begin{equation}\label{generalisedkomarr}
\mathcal{U}=\nabla_{\alpha}\xi^{[\mu}P^{\nu]\alpha}ds_{\mu\nu},
\end{equation}
where $ds_{\mu\nu}$ denotes the surface elements for a given background and
is the standard basis for $n-2$-forms over the manifold $M$ ($n=\mathrm{dim}(M)$).

\section{Thermodynamics of Kerr black hole} 
 A solution to the Einstein field equations describing  rotating black holes was discovered by Roy Kerr. This is a solution that only describes a rotating black hole without charge. Indeed, there is a solution for charged black holes (\textit{i.e.} satisfies Einstein-Maxwell equations) known as Kerr-Newman. Kerr metric can be written in number of ways and in this paper we are going to use the Boyer-Lindquist coordinate. The metric is given by \cite{eric} 
 \begin{align}\label{metric}
ds^{2}&=-(1-\frac{2Mr}{\rho^{2}})dt^{2}-\frac{4Mar\sin^{2}\theta}{\rho^{2}}dtd\phi+\frac{\Sigma}{\rho^{2}}\sin^{2}\theta d\phi^{2}+\frac{\rho^{2}}{\Delta}dr^{2}+\rho^{2}d\theta^{2},
\end{align}
where, 
\begin{eqnarray}
\rho^2=r^{2}+a^2 \cos^2 \theta, \quad \Delta=r^{2}-2Mr+a^{2},\quad \Sigma=(r^{2}+a^2 )^{2}-a^{2}\Delta\sin^{2}\theta.
\end{eqnarray}
The metric is singular at $\rho^2=0$. This singularity is real\footnote{This is different
than the singularity at $\Delta=0$ which is a coordinate singularity.} and can be checked via Kretschmann scalar\footnote{The Kretschmann scalar for Kerr metric is given by: $R^{\alpha\beta\gamma\delta}R_{\alpha\beta\gamma\delta}=\frac{48M^{2}(r^{2}-a^{2}\cos^{2}\theta)(\rho^{4}-16a^{2}r^{2}\cos^{2}\theta)}{\rho^{12}}$.} \footnote{We shall note that scalar curvature,
$R$, and Ricci tensor, $R_{\mu\nu}$ are vanishing for the Kerr metric and
only some components of the Riemann curvature are non-vanishing.}. The above metric has two horizons $r_{\pm}=m\pm\sqrt{m^{2}-a^{2}}$. Furthermore, $a^2 \leq m^2$  is a length scale. Let us define the vector:
 \begin{equation}\label{comb}
\xi^{\alpha}=t^{\alpha}+\Omega\phi^{\alpha}.
\end{equation}
This vector is null at the event horizon. It is tangent to the horizon's null generators, which wrap around the horizon with angular velocity $\Omega$. Vector $\xi^{\alpha}$ is a Killing vector since it is equal to sum of two Killing vectors. After all, the event horizon of the Kerr metric is a Killing horizon.  Using Eqs. (\ref{generalisedkomarr}) and (\ref{comb}) we can define \textit{Komar integrals} describing the energy and the angular momentum of the Kerr black hole as, 
\begin{equation}
\mathcal{E}=-\frac{1}{8\pi}\lim_{S_{t}\rightarrow \infty}\oint_{S_{t}}\nabla_{\lambda}P^{\alpha{\lambda}}\xi^{\beta}_{(t)}ds_{\alpha\beta},
\end{equation}
\begin{equation}
\mathcal{J}=\frac{1}{16\pi}\lim_{S_{t}\rightarrow \infty}\oint_{S_{t}}\nabla_{\lambda}P^{\alpha{\lambda}}\xi^{\beta}_{(\phi)}ds_{\alpha\beta},
\end{equation}
where the integral is over $S_{t}$, which is a closed two-surfaces\footnote{Note
that we can write $\lim_{S_{t}\rightarrow \infty}\oint_{S_{t}}$ as simply
$\oint_{\mathcal{H}}$ where $\mathcal{H}$ is a two dimensional cross section
of the event horizon.}. This is
because the equilibrium state version of the first law of black holes is
a balance sheet of energy between two stationary black holes.
We shall note that $S_{t}$ is an $n-2$ surface. In above definitions $\xi^{\beta}_{(t)}$
is the space-time's time-like Killing vector and $\xi^{\beta}_{(\phi)}$ is
the rotational Killing vector and they both satisfy the Killing's equation,
$\xi_{\alpha;\beta}+\xi_{\beta;\alpha}=0$. Moreover, the sign difference
in two definition has its root in the signature of the metric. In this
paper we are using $(-,+,+,+)$ signature. The surface element is also given
by, 
 \begin{equation}
ds_{\alpha\beta}=-2n_{[\alpha}r_{\beta]}\sqrt{\sigma}d\theta
d\phi,
\end{equation}
where $n_{\alpha}$ and $r_{\alpha}$ are the time-like (\textit{i.e.} $n_{\alpha}n^{\alpha}=-1$) and space-like (\textit{i.e.} $r_{\alpha}r^{\alpha}=1$) normals to $S_{t}$. For Kerr metric in Eq. (\ref{metric})
the normal vectors are defined as: 
\begin{equation}
n_\alpha=(-\frac{1}{\sqrt{-g^{tt}}},0,0,0)=(-\sqrt{\frac{\rho^{2}\Delta}{\Sigma}},0,0,0),
\end{equation}
\begin{equation}
r_\beta=(0,\frac{1}{\sqrt{g^{rr}}},0,0)=(0,\sqrt{\frac{\rho^{2}}{\Delta}},0,0).
\end{equation}
Furthermore, the  two dimensional cross section
of the event horizon described by $t=$constant and also $r=r_+$ (\textit{i.e.} constant), hence, from metric in Eq. (\ref{metric}) we can extract the induced metric as: 
\begin{equation}
\sigma_{AB}d\theta^{A}d\theta^{B}=\rho^{2}d\theta^{2}+\frac{\Sigma}{\rho^{2}}\sin^{2}\theta d\phi^{2}.
\end{equation}
Thus we can write, 
\begin{equation}\label{detinduced}
\sqrt{\sigma}=\sqrt{\Sigma}\sin\theta d\theta d\phi.
\end{equation}

\textit{First law} of black hole thermodynamics states that when a stationary
black hole at manifold $\mathcal{M}$ is perturbed slightly to $\mathcal{M}+\delta\mathcal{M}$,
 the difference in the energy, $\mathcal{E}$, angular momentum, $\mathcal{J}_{a}$,
and area, $\mathcal{A}$, of the black hole are related by:
\begin{eqnarray}\label{mainentropy}
\delta \mathcal{E}=\Omega^{a}\delta \mathcal{J}_{a}+\frac{\kappa}{8\pi}\delta
\mathcal{A}=\Omega^{a}\delta \mathcal{J}_{a}+\frac{\kappa}{2\pi}\delta
\mathcal{\mathcal{S}},
\end{eqnarray}
where  $\Omega^{a}$ are the angular velocities at the horizon. We shall note that $\mathcal{S}$ is the associated entropy.
$\kappa$ denotes the \textit{surface gravity} of the Killing horizon 
and for the metric given in Eq. (\ref{metric}) the surface gravity is given
by
\be\label{surfacegravitykerr}
\ka=\frac{\sqrt{m^2-a^2}}{2mr_{+}} \,.
\ee
 The surface area \cite{eric} of the black hole is given by\footnote{We shall note that $\mathcal{S}=\mathcal{A}/4$
(with $G=1)$ denotes the Bekenstein-Hawking entropy.
}: 
\begin{equation}
\mathcal{A}=\oint_{\cH}\sqrt{\sa}d^2 \theta,
\end{equation}
where $d^2 \theta=d\theta d\phi$. Now by using Eq. (\ref{detinduced}), the surface
area can be obtained as, \begin{equation}
\mathcal{A}=\oint_{\cH}\sqrt{\sa}d^2 \theta=\int_{0}^{\pi} \sin (\theta)d\theta\int_{0}^{2\pi}d\phi(r^{2}_{+}+a^{2})=4\pi(r^{2}_{+}+a^{2}).
\, 
\end{equation}

\section{Modified gravity}
Modified theories of gravity were proposed as an attempt to describe some of the phenomena that Einstein's theory of general relativity can not address. Examples of these phenomena can vary from explaining the singularity to the dark energy. In this section, we obtain the entropy of the Kerr black hole for number of these theories.   
\subsection{Einstein-Hilbert action}
As a warm up exercise let us start the calculation  for  the most well knows case, where the action is given by: \textbf{}
\begin{equation}
S_{EH}=\int d^{4}x\sqrt{-g} R,
\end{equation} 
for this case, as shown in footnote \ref{ehpotential}, the Komar integrals can be found explicitly as \cite{Fatibene:1998rq}, 
\begin{eqnarray}
\mathcal{E}&=&-\frac{1}{8\pi}\oint_{\mathcal{H}}\nabla^{\alpha}t^{\beta}ds_{\alpha\beta}\nonumber\\
&=&-\frac{1}{8\pi}\int^{2\pi}_{0} d\phi\int^{\pi}_{0} d\theta\nonumber\\&&\Bigg(\frac{1}{2}
\sin (\theta ) \left(a^2 \cos (2 \theta )+a^2+2 r^2\right)\frac{8 m \left(a^2+r^2\right)
\left(a^2 \cos (2 \theta )+a^2-2 r^2\right)}{\left(a^2
\cos (2 \theta )+a^2+2 r^2\right)^3}\Bigg)=m. 
\end{eqnarray}
We took $\xi^{\alpha}=t^{\alpha}$, where $t^{\alpha}=\frac{\partial
x^{\alpha}}{\partial t}$; $x^{\alpha}$ are the space-time coordinates.
So, for instance, $g_{\mu \nu}\xi^{\mu}\xi^{\nu}=g_{\mu \nu}t^{\mu}t^{\nu}=g_{tt}$,
that is after the contraction of the metric with two Killing vectors, one
is left with the $tt$ component of the metric.
In similar manner, we can calculate the angular momentum as, 
\begin{eqnarray}
\mathcal{J}&=&\frac{1}{16\pi}\oint_{\mathcal{H}}\nabla^{\alpha}\phi^{\beta}ds_{\alpha\beta}\nonumber\\
&=&\frac{1}{16\pi}\int^{2\pi}_{0} d\phi\int^{\pi}_{0} d\theta\nonumber\\&&\Bigg(\frac{1}{2}
\sin (\theta ) \left(a^2 \cos (2 \theta )+a^2+2 r^2\right)\nonumber\\&\times&\frac{-8
a m \sin ^2(\theta ) \left(a^4-3 a^2 r^2+a^2 (a-r) (a+r) \cos (2 \theta )-6
r^4\right)}{\left(a^2 \cos (2 \theta )+a^2+2 r^2\right)^3}\Bigg)=ma.\end{eqnarray}
Now given Eq. (\ref{mainentropy}), we have, 
\begin{equation}
\frac{\kappa}{2\pi}\delta
\mathcal{\mathcal{S}}=\delta \mathcal{E}-\Omega^{a}\delta \mathcal{J}_{a}=(1-\Omega a)^{}\delta m-\Omega^{}m^{}\delta a.
\end{equation}
By recalling the surface gravity from Eq. (\ref{surfacegravitykerr})
we have, 
\be
\mathcal{S}=2\pi m r_{+}. 
\ee
which is a well known result. 
\subsection{$f(R)$ theories of gravity} 
There are numerous ways to modify the Einstein theory of general relativity, one of which is going to higher order curvatures. A class of theories which attracted attention in recent years is the $f(R)$ theory of gravity \cite{Sotiriou:2008rp}. This type of theories can be seen as simply the series expansion of the scalar curvature, $R$, and one of the very important features of them is that they can avoid Ostrogradski instability \cite{Woodard:2006nt}.
The action of this gravitational theory is generally given by:
\begin{equation}
S_{f(R)}=\int d^{4}x\sqrt{-g} f(R),
\end{equation}
where $f(R)$ is the function of scalar curvature and it can be of any order. In this case the Komar potential can be obtained by,
 
\begin{equation}
P^{\alpha\beta}_{f(R)}=\frac{\delta \mathcal{L}}{\delta R_{\alpha\beta}}=\frac{\delta f(R)}{\delta R}\frac{\delta R}{\delta R_{\alpha\beta}}=f'(R)\sqrt{-g}g^{\alpha\beta},
\end{equation}
and thus: 
\begin{equation}
\mathcal{U}_{f(R)}=f'(R)\sqrt{-g}\nabla_{\alpha}\xi^{[\mu}g^{\nu]\alpha}ds_{\mu\nu}.
\end{equation}
This results in modification of the energy and angular momentum as, 
\begin{equation}
\mathcal{E}_{f(R)}=-\frac{1}{8\pi}\oint_{\mathcal{H}}f'(R)\nabla^{\alpha}t^{\beta}ds_{\alpha\beta}=f'(R)m,
\end{equation}
and
\begin{equation}
\mathcal{J}_{f(R)}=\frac{1}{16\pi}\oint_{\mathcal{H}}f'(R)\nabla^{\alpha}\phi^{\beta}ds_{\alpha\beta}=f'(R)ma.
\end{equation}
We know that the $f(R)$ theory of gravity is essentially the power expansion in the scalar curvature, \begin{equation}
f(R)=R+R^{2}+R^{3}+\cdots+R^{n},
\end{equation}
and thus, 
\begin{equation}\label{diff}
f'(R)=1+2R^{}+3R^{2}+\cdots+nR^{n-1}.
\end{equation}
As a result, the entropy of $f(R)$ theory of gravity is given only by the Einstein Hilbert contribution, 
\be
\mathcal{S}_{f(R)}=\mathcal{S}_{EH}=2\pi m r_{+}. 
\ee
This is due to fact that the scalar curvature, $R$, is vanishing for the Kerr metric given in Eq. (\ref{metric}) and so only the leading term in Eq. (\ref{diff}) will be accountable. 
\subsection{$f(R,R_{\mu\nu})$}
After considering the $f(R)$ theories of gravity, it is natural to think about the more general form of gravitational modification. In this case: $f(R,R_{\mu\nu}),$  the action would contain terms like $R_{\mu\nu}R^{\mu\nu}$, $R^{\mu\alpha}R^{ \ \nu}_{\alpha}R_{\nu\mu}$ and so on. Let us take a specific example of, 
\begin{equation}
S_{R_{\mu\nu}}=\int d^{4}x\sqrt{-g} (^{}R+\lambda_{1}R_{\mu\nu}R^{\mu\nu}+\lambda_{2}R^{\mu\lambda}R^{
\ \nu}_{\lambda}R_{\nu\mu}),
\end{equation}
where $\lambda_{1}$ and $\lambda_{2}$ are coefficients of appropriate dimension (\textit{i.e.} mass dimension $L^2$ and $L^4$ respectively where $L$ denotes length).  The momenta would then be obtained as, 
\begin{align}
P^{\al \bt}_{R_{\mu\nu}}=\frac{\p\mathcal{L}}{\p R_{\al \bt}}=\sqrt{-g}(^{}g^{\al \bt}+2\lambda_{1}R^{\al \bt}+3\lambda_{2}R^{\bt \lambda}R^{\al}_{\ \lambda}).
\end{align}
As before, the only contribution comes from the EH term since the Ricci tensor is vanishing for the Kerr metric given in  Eq.
(\ref{metric}). So, without proceeding further, we can conclude that in this case the entropy is given by the area law only and  with no correction.  
\subsection{Higher derivative gravity}

Another class of modified theories of gravity are the higher derivative theories. We shall denote the action by $S(g,R,\nabla R,\nabla R_{\mu\nu},\cdots)$. In this class, there are covariant derivatives acting on the curvatures. Moreover, there are theories that contain inverse derivatives acting on the curvatures \cite{Conroy:2014eja}. These are known as non-local theories of gravity and we do not wish to consider them in this paper. 

 A well established class of higher derivative theory of gravity is given by \cite{Biswas:2005qr} where the action contains infinite derivatives acting on the curvatures. It has been shown that having infinite derivatives can cure the singularity problem \cite{Biswas:2011ar}. This is achieved by replacing the singularity with a bounce. Moreover, this class of theory preserves the ghost freedom. This is of a very special importance, since in other classes of modified gravity, deviating from the EH term and going to higher order curvature terms means one will have to face the ghost states.  Having infinite number of derivatives makes it extremely difficult to find a metric solution which satisfies the equations of motion. Moreover,  infinite derivative theory is associated with singularity freedom and Kerr metric is a singular one. As a result, in this paper we wish to consider a finite derivative example as a matter of illustration,  let us define the Lagrangian of the form: 
\be\label{HD}
\cL_{HD}=\sqrt{-g} \LT R+\sum_{n=1}^{m_1}R \bar{\Box}^{n}R+\sum_{n=1}^{m_2}R_{\mu
\nu} \bar{\Box}^{n}R^{\mu \nu} \RT \,,
\ee
where $\Box=g^{\mu\nu}\nabla_{\mu}\nabla_{\nu}$ is d'Alembertian operator and $\bar{\Box}=\Box/M^{2}$ to ensure the correct dimensionality. We also note that $m_1$ and $m_2$ are some finite number. In this case, we have the Lagrange momenta as, 
\begin{align}
P^{\al \bt}_{HD}&=\frac{\p \cL}{\p R_{\al \bt}} 
=\sqrt{-g} \LT g^{\al \bt}+2g^{\al \bt}\sum_{n=1}^{m_1} \bar{\Box}^{n}R+2\sum_{n=1}^{m_2}
\bar{\Box}^{n}R^{\al \bt} \RT  
 .
\end{align}
As mentioned previously for the Kerr metric:  $R=R^{\al \bt}=0$, this is to conclude that the only non-vanishing term which will contribute to the entropy will be the first term, in the above equation, which corresponds to the EH term in the action given in Eq. (\ref{HD}).

\section{Conclusion}
In this paper we proposed how to calculate the entropy for the Kerr background in various  examples. It has been shown that deviating from the EH\ gravity up to Ricci tensor will have no effect in the amount of entropy, and the entropy is given solely by the area law. This is because the scalar curvature and Ricci tensor are vanishing for the Kerr metric given in Eq.
(\ref{metric}). 

However, if the action contains Riemann tensors (such as $R_{\alpha\beta\gamma\delta}R^{\alpha\beta\gamma\delta},R_{\alpha\beta\gamma\delta}\Box R^{\alpha\beta\gamma\delta}$, and other possible combinations of Riemann tensor)  then there will be a modification to the entropy. Calculating the conserved charge and thus the entropy on the Kerr background for those theories containing Riemann tensor is a hard task. This is due to the fact that the Kerr metric as given in Eq.
(\ref{metric}) contains many non-vanishing Riemann tensor components and thus we shall leave those cases for future studies. Moreover, the methodology to obtain the entropy where we have Riemann tensor is slightly different \cite{Peng:2014gha}. 
\section*{Acknowledgment}
S.T is supported by a scholarship from the Onassis Foundation. 

\appendix

\section{Conserved current for Einstein-Hilbert gravity}\label{ehcons}
Given the EH action to be of the form, 
\begin{equation}
S_{EH}=\int d^{4}x\sqrt{-g} R.
\end{equation}
we can imply the variation principle infinitesimally by writing,
\begin{eqnarray}\label{ehvar}
\delta_{\xi}S_{EH}=\int d^{4}x\delta_{\xi}(\sqrt{g} R)
=\int d^{4}x\sqrt{g}\Big(G_{\mu\nu}\delta_{\xi}g^{\mu\nu}+g^{\mu\nu}\delta_{\xi}(R_{\mu\nu})\Big)
=\int d^{4}x\sqrt{g}\nabla_\alpha(\xi^\alpha R)=0
\end{eqnarray}
where $G_{\mu\nu}$ is the Einstein tensor and given by $G_{\mu\nu}=R_{\mu\nu}-\frac{1}{2}g_{\mu\nu}R$.
The term involving the Einstein tensor can be expanded further as, 
\begin{eqnarray}\label{term1}
G_{\mu\nu}\delta_{\xi}g^{\mu\nu}=G_{\mu\nu}(\nabla^\mu\xi^\nu+\nabla^\nu\xi^\mu)=2G_{\mu\nu}\nabla^\mu\xi^\nu=\nabla_\mu(-2R^{\mu}_{\nu}+\delta^{\mu}_{\nu}R)\xi^\nu
\end{eqnarray}
where we used Eq. (\ref{metricvar}) and performed integration by parts. Then
we move on to the next term and expand it as, 
\begin{equation}\label{term2}
g^{\mu\nu}\delta_{\xi} R_{\mu\nu}=(\nabla^{\mu}\nabla^{\nu}-g^{\mu\nu}\Box)\delta_{\xi}
g_{\mu\nu}=\nabla_{\lambda}\Big((g^{\lambda\alpha}g^{\nu\beta}-g^{\lambda\nu}g^{\alpha\beta})\nabla_{\nu}(\nabla_\alpha\xi_\beta+\nabla_\beta\xi_\alpha)\Big)
\end{equation}
by substituting Eq's. (\ref{term1}) and (\ref{term2}) into (\ref{ehvar})
we obtain, 
\begin{eqnarray}
\delta_{\xi}S_{EH}
=\int d^{4}x\sqrt{-g}\nabla_\mu\Big(-2R^{\mu}_{\nu}\xi^\nu+(g^{\mu\alpha}g^{\nu\beta}-g^{\mu\nu}g^{\alpha\beta})\nabla_{\nu}(\nabla_\alpha\xi_\beta+\nabla_\beta\xi_\alpha)\Big)=0
\end{eqnarray}
and hence for any vector field $\xi^{\mu}$ one obtains the conserved N\"oether
current, 
\begin{equation}\label{noethercurrent}
J^{\mu}(\xi)=R^{\mu}_{\nu}\xi^\nu+\frac{1}{2}(g^{\mu\alpha}g^{\nu\beta}-g^{\mu\nu}g^{\alpha\beta})\nabla_{\nu}(\nabla_\alpha\xi_\beta+\nabla_\beta\xi_\alpha)\equiv\nabla_\nu(\nabla^{[\mu}\xi^{\nu]})
\end{equation}

\section{Generalised Komar current}\label{generalisedkomar}
It can be shown that the N\"oether current that was obtained in Eq. (\ref{noethercurrent})
is identical to generalised Komar current via
\begin{eqnarray}
J^{\mu}(\xi)&=&\frac{1}{2}\nabla_\nu(\nabla^{\mu}\xi^{\nu}-\nabla^{\nu}\xi^{\mu})=\nabla_\nu\nabla^{\mu}\xi^{\nu}-\frac{1}{2}\nabla_\nu(\nabla^{\nu}\xi^{\mu}+\nabla^{\mu}\xi^{\nu})\nonumber\\
&=&[\nabla^{\nu},\nabla^\mu]\xi_\nu+\nabla^\mu(\nabla_\nu\xi^\nu)-\frac{1}{2}\nabla_\nu(\nabla^{\nu}\xi^{\mu}+\nabla^{\mu}\xi^{\nu})\nonumber\\
&=&R^{\mu}_{\nu}\xi^\nu+\frac{1}{2}(g^{\mu\alpha}g^{\nu\beta}-g^{\mu\nu}g^{\alpha\beta})\nabla_{\nu}(\nabla_\alpha\xi_\beta+\nabla_\nu\xi_\beta)
\end{eqnarray}
where we\ used: $[\nabla^{\nu},\nabla^\mu]\xi_\nu=R^{\mu}_{\lambda\mu\nu}\xi^{\lambda}=R_{\lambda\nu}\xi^{\lambda}$.



\end{document}